\documentclass[
    ,final            
  ]
  {aipproc}

\layoutstyle{6x9}

\begin{document}

\newcommand{\MET}       {$E_T\hspace{-2.4ex}/\hspace{1.2ex}~$}
\newcommand{\pt}     {\ \mathrm{p_T}}
\newcommand{\GeVcc}     {\ \mathrm{GeV/c^2}}
\newcommand{\GeVc}     {\ \mathrm{GeV/c}}
\newcommand{\GeV}       {\ \mathrm{GeV}}
\newcommand{\invpb}     {~\mathrm{pb}^{-1}}
\newcommand{\invfb}     {~\mathrm{fb}^{-1}}

\newcommand{\ETM}       {E_T\hspace{-2.4ex}/\hspace{1.2ex}}
\newcommand{\HT}        {$H_T$}
\newcommand{\gl}        {\tilde{g}}
\newcommand{\sq}        {\tilde{q}}
\newcommand{\sqb}       {\bar{\tilde{q}}}
\newcommand{\qb}        {\bar{q}}
\newcommand{\mathgl}    {\mathrm{\tilde{g}}}
\newcommand{\mathsq}    {\mathrm{\tilde{q}}}

\title{Searches for Squarks and Gluinos at CDF and D\O~ Detectors}


\author{Xavier Portell\footnote{Speaker, on behalf of the CDF and D\O
Collaborations.}}{
  address={IFAE, Barcelona, Spain\\
        portell@fnal.gov}
}

\begin{abstract}
 This contribution reports on preliminary measurements on searches for squarks
and gluinos at CDF and D\O~ detectors in $p\bar{p}$ collisions at
$\sqrt{s}=1.96$~TeV. The analyses are performed using event topologies with multiple
jets and large missing energy in the final state. The mSUGRA scenario and
R-parity conservation is assumed. No excess with respect to the Standard Model predictions is
observed and new limits on the gluino and squark masses are extracted.
\end{abstract}

\maketitle


\section{INTRODUCTION}

Supersymmetry (SUSY)~\cite{susy} is an extension of the Standard
Model (SM) that naturally solves the hierarchy problem and, at the same time, provides a good
candidate for cold dark matter in the universe. SUSY introduces a fermion-boson symmetry predicting, for every
particle in the SM, the existence of a super-partner. In mSUGRA~\cite{msugra}, the spectrum of particles is defined by five different parameters: $M_0$, $M_{1/2}$, $A_0$, $\tan\beta$ and $sign(\mu)$). When R-Parity (R$_P$) is conserved, the new supersymmetric particles would be produced in pairs and ultimately decay into the lightest supersymmetric
particle (LSP), which is stable and escapes the experimental apparatus
undetected. This leads to event topologies characterized by the presence of multiple jets, from the different decays of heavy sparticles, and large missing transverse energy (\MET) in the final state, and have been investigated by the CDF~\cite{CDF} and D\O~\cite{D0} detectors.

\section{Background Processes}
The supersymmetric signal must be extracted from large background contributions. QCD multijet processes, where the \MET is originated by jets reconstructed in partially instrumented regions of the detector, constitutes one of the most important backgrounds, together with the production of $Z$ and $W$ bosons in association with jets, where the missing transverse energy is originated by the presence of neutrinos in the final state or the misidentification of jets. In particular, $Z\to\nu\bar\nu$ + jets constitutes an irreducible background to the supersymmetric signature. In addition, $WW$ and $t\bar t$ production, among others, constitute significant background processes that must be taken into account. D\O~ and CDF employed two different methods to estimate the dominant QCD background. D\O~ relies on a fit to an exponential falling shape, based on QCD data at low \MET, and extrapolates it to the very high \MET region. CDF decided to generate massive MC samples to take into account non-gaussian tails in the \MET distribution and then, used the data to check the MC absolute yields. For the rest of backgrounds, both experiments extracted predictions from Monte Carlo generators (ALPGEN~\cite{alpgen} in the case of Z/W+jets and PYTHIA~\cite{pythia} for WW and $t\bar t$ production) normalized to NLO predictions.

\section{Experimental Strategies and Results}
  A mSUGRA scenario with $A_0=0$, sign($\mu$) = -1  and $\tan\beta=3$ ($\tan\beta = 5$) 
in the case of D\O~ (CDF) have been assumed. Both experiments applied 
similar pre-selection cuts to remove cosmics and beam-related backgrounds. 

\begin{figure}
  \includegraphics[height=.3\textheight]{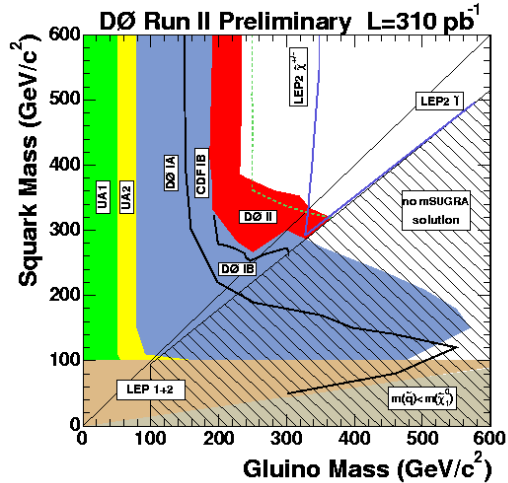}
  \caption{95\% CL exclusion regions in the gluino-squark mass plane, where mSUGRA with $\tan\beta=3, A_0=0$ and sign($\mu$)<0 is assumed.The different shadow bands denote the result  of different experiments. The hashed region contains no mSUGRA solution and the dashed line is the expected limit.}
  \label{D0limits}
\end{figure}

In the case of D\O~ the analysis was based on $310\invpb$ and the 
cuts were optimized in three different regions of the gluino-squark mass plane. 
When $M_\mathgl > M_\mathsq$, the squark production is enhanced and the final state signature is characterized by 
dijet events and large missing transverse energy, 
since the produced squarks tend to decay into a jet and a LSP. 
When $M_\mathgl < M_\mathsq$, the gluino production is more important and the final-state topologies 
are dominated by the presence of at least four jets. 
Finally, when $M_\mathgl \sim M_\mathsq$  the analysis requires at least 3 jets. To improve the signal significance, the scalar sum of transverse energies is optimized to be above 250-375 GeV and the \MET to be above 75-175 GeV.
In all considered topologies the 
observed number of events is in good agreement with SM predictions. The resulting 
limits in the gluino-squark mass plane can be seen in Figure~\ref{D0limits}.

\begin{figure}
  \includegraphics[height=.3\textheight]{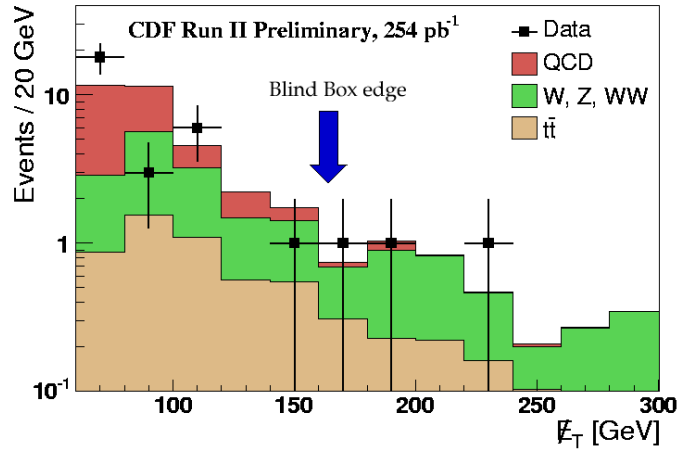}
  \caption{\MET distribution for CDF after the rest of the cuts have been applied. The arrow indicates de boundary 
of the defined signal region for the analysis. The points are the data and the histograms are the 
different backgrounds, where each includes the previous one and QCD denotes the total SM background.}
\label{CDF_MET}
\end{figure}

In the case of CDF, a {\it blind analysis technique} based on the first $254\invpb$ of data was carried out, and the analysis cuts were optimized in a region were
the mSUGRA signal was maximal (for $M_\mathgl \sim M_\mathsq ~ 340\GeVcc$) . Events were required to have  
at least three jets with a scalar sum of transverse energies above 350 GeV and \MET above 165 GeV. In this case, (see Figure~\ref{CDF_MET}) 
the observed number of events was also found in good agreement with the SM predictions.    

\section{Summary and Conclusions}

CDF and D\O~ found no evidence for the production of gluinos and squarks in events with 
multiple jets and large missing transverse energy in the final state based on 
about $300\invpb$ of data. As result,  Run I exclusion limits in the gluino-squark mass plane
have been significantly expanded.


\begin{theacknowledgments}
I would like thank the organizers for the kind invitation 
and their great hospitality during the conference. 
\end{theacknowledgments}

\end{document}